\documentclass[letterpaper]{article}
\usepackage[preprint]{aaai2027}
\usepackage[hyphens]{url}
\usepackage{graphicx}
\usepackage{natbib}
\usepackage{caption}
\usepackage{booktabs}
\usepackage{amsmath}
\usepackage{tikz}
\usetikzlibrary{arrows.meta,positioning}

\title{Matryoshka Hash Representations for Model-Aware Compact Semantic Retrieval}
\author{
    Peichun Hua,
    Yunming Xiao\corresponding
}
\affiliations{
    The Chinese University of Hong Kong, Shenzhen\\
    peichunhua@link.cuhk.edu.cn, yunmingxiao@cuhk.edu.cn
}

\begin{document}
\maketitle

\begin{abstract}
Retrieval-augmented generation (RAG) depends on dense retrieval: each document is stored as a learned vector, and a query is answered by finding its nearest neighbors in that vector space. Keeping one full-precision vector per document is the dominant index cost at corpus scale, so retrieval systems replace each vector with a short code of a few bytes, a step called quantization. Standard quantizers such as product quantization (PQ) pick the code that reconstructs the original vector most closely. Reconstruction is only a proxy for retrieval: at a few bytes per document, a \emph{model-aware} code trained on the retriever's ranking signal uses the budget better.

A single code is even more useful if it serves several byte budgets at once: when its short prefixes are each directly searchable, a deployment can set its efficiency--quality operating point without re-encoding the corpus.
But training all prefixes under one objective makes the early bits a compromise across budgets: short codes improve while the full-width code degrades.
Binary codes sharpen the conflict, since a gradient step either flips a stored bit or leaves it unchanged.
We introduce Matryoshka Hash Representations (MHR), a two-stage procedure that separates full-width training from prefix organization. MHR first learns a longer binary code, then freezes the model and trains additional zero-initialized residual code adaptors for directly searchable prefixes.
Documents are stored at one bit per coordinate, queries keep continuous logits as in PQ, and search runs on FAISS FastScan. Trained on MS~MARCO and zero-shot transferred to seven BEIR datasets, one MHR code reaches .502 NDCG@10 at 16 bytes and .556 at 32 bytes, against .404 and .524 for the best compact baselines; the gains hold across training seeds, on all 13 public BEIR datasets, and against off-the-shelf binary Matryoshka embeddings with stronger backbones. The same code also strengthens two common pipelines: shortlisting candidates for full-precision reranking, and pruning a low-storage graph index such as LEANN.
\end{abstract}

\section{Introduction}

Dense retrieval is the first stage of many text systems. It grounds retrieval-augmented generation (RAG), which answers a query by retrieving relevant documents and conditioning a language model on them rather than on its parameters alone~\cite{lewis2020rag}, and it also drives open-domain question answering~\cite{karpukhin2020dpr}, large-scale semantic search~\cite{douze2026faiss,hua2026semantics}, candidate generation in recommender systems~\cite{covington2016youtube,yi2019sampling}, and evidence retrieval for fact verification~\cite{wadden2020factfiction}. Dense implementations of these applications map queries and documents into a shared vector space and reduce retrieval to nearest-neighbor search. Storing one full-precision vector per document then becomes the dominant index cost as the corpus grows.
Vector compression techniques such as product quantization (PQ) and optimized product quantization (OPQ) reduce this cost by compressing vectors after encoder training~\cite{jegou2011ivfpq,ge2013opq}.
These quantizers are \emph{model-agnostic}: they fit a code by minimizing reconstruction error and preserving the geometry of the original embeddings. They spend much of their capacity on variation that leaves the top-ranked documents unchanged.
A \emph{model-aware} code is instead trained on the retrieval model's own ranking signal through relevance ordering and anchoring on signals from the full-precision embeddings.
Model-awareness comes from the objective, so a code trained to rank on top of a frozen encoder still qualifies. It matters most at a few bytes per document, where every bit counts, and it separates PQ/OPQ, ITQ~\cite{gong2011iterative}, RaBitQ~\cite{gao2024rabitq}, and LSH-style codes~\cite{charikar2002similarity,ji2012superbit} from ranking-trained discrete retrievers.

Nested coding, introduced as Matryoshka representation learning~\cite{kusupati2022mrl}, lets one stored representation serve every prefix, so a deployment can select its byte budget without maintaining an index per width. Nested codes also enable adaptive retrieval: a short prefix filters candidates before a longer one refines the shortlist \cite{rege2023adanns}. This is especially useful for binary hashing, where re-encoding requires model access and storage already approaches tens of bytes per document.

However, nesting has a cost. Because each prefix must rank well on its own, the training objective places a ranking loss at every width, and the early bits fall under all of these losses at once. The short prefixes then improve at the full-width code's expense: a \emph{full-width--prefix trade-off}. In real-valued MRL, this is softened by fitting the prefixes after training or one width at a time \cite{yoon2024matryoshka,zhang2025smec}. Binary codes sharpen this trade-off. A continuous coordinate moves a little under each gradient step, but a stored bit is that coordinate's sign: a step either flips the bit or leaves it untouched, so the competing widths flip early bits back and forth instead of settling.

In this paper, we propose Matryoshka Hash Representations (MHR), which separates the two goals that direct nesting entangles. We specifically target the low-storage regime of 64--256 bits per document, as sought by recent indexing systems such as LEANN~\cite{wang2025leann}.
MHR first adapts a pretrained dual encoder and learns a 256-bit hash code on MS~MARCO with LoRA~\cite{hu2022lora}, compressing 768-dimensional floats by 96$\times$. It then freezes the encoder and trains two additional residual adaptors over the fixed logits.
We adopt a full-width rank distillation that anchors the 256-bit code, while prefix ranking losses organize the adapted logits into nested 64-, 128-, and 256-bit codes that do not entangle with one another. 
At deployment, documents are binarized and queries keep their continuous logits.

MHR needs no graph topology~\cite{subramanya2019diskann,malkov2018hnsw}: a flat index scores every code, and IVF selects coarse lists and ranks their MHR-coded postings. We further show that the same code supports graph indexes that prune neighbor expansions~\cite{wang2025leann} and coarse-to-fine pipelines that shortlist candidates before reranking them with full-precision embeddings. Across these index types we ask one question: can a binary code be trained to produce short, searchable prefixes without losing full-width quality?

Our contributions are:
\begin{itemize}
    \item We identify a full-width--prefix trade-off in direct nested binary training and introduce a two-stage procedure that decouples the two goals. Our MHR framework first learns a strong full-width source code, then trains a residual adaptor cascade over the frozen code with prefix supervision, raw-query scoring matched to deployment, and a full-width rank anchor to obtain a series of code widths.
    \item We align the compact score with deployment requirements, using hard document codes and continuous queries. Our FAISS-native flat and IVF kernels scan faster than PQ indexes of similar size.
    \item On seven BEIR datasets, MHR leads every compact baseline in macro NDCG@10 and Recall@100 at all budgets. Ablations support each design choice, and MHR is a stronger drop-in replacement for PQ in flat, IVF, graph, and reranking pipelines.
\end{itemize}

\section{Related Work}

\paragraph{Matryoshka text embedding.} MRL jointly supervises real-valued prefixes~\cite{kusupati2022mrl}, and AdANNS selects among them within an ANN pipeline~\cite{rege2023adanns}. Matryoshka-Adaptor adds truncatability without retraining the encoder~\cite{yoon2024matryoshka}, while SMEC sequentializes compression to mitigate width-dependent gradient variance~\cite{zhang2025smec}. Binary MRL truncates and signs such embeddings, as offered for nomic-embed-v1.5 and mxbai-embed \cite{nussbaum2025nomic,nomic2024v15,koenig2024binarymrl}, and Matryoshka Quantization nests integer precisions of model weights \cite{nair2025matquant}. These methods keep a real-valued training target; MHR instead trains hard one-bit coordinates for nested, searchable low-bit prefixes.

\paragraph{Multi-length supervised hashing.} JMH~\cite{liu2019jmh} jointly learns several code lengths, MAH~\cite{luo2021mah} transfers information among predefined low-bit codes, and the Nested Hash Layer~\cite{he2024nhl} distills longer codes into nested prefixes. These methods target \textit{supervised} image retrieval, where class or pair labels define semantic neighborhoods. MHR instead learns fine-grained query–document ordering and transfers one nested text code to corpora \textit{without} target relevance labels.

\paragraph{In-memory compact semantic retrieval.} Post-hoc methods include PQ and OPQ \cite{jegou2011ivfpq,ge2013opq}, the bounded binary estimator and per-vector corrections of RaBitQ and its multi-bit extension \cite{gao2024rabitq,gao2025extendedrabitq}, and the data-dependent isolation partitions of IKE \cite{zhang2026ike}. Learned alternatives include mutual-information document hashing \cite{ou2021dhim}, binary passage indexes \cite{yamada2021bpr,gan2023bebr}, and jointly optimized discrete PQ retrievers such as JPQ and RepCONC \cite{zhan2021jpq,zhan2022repconc}; target-specific synthetic data have also been used to adapt discrete codes under domain shift \cite{thakur2023domainhash}. MHR learns a discrete representation once on source data and retrieves on unseen corpora at low storage and scan cost.

\section{Matryoshka Hash Representations}

\begin{figure*}[t]
\centering
\includegraphics[width=\textwidth]{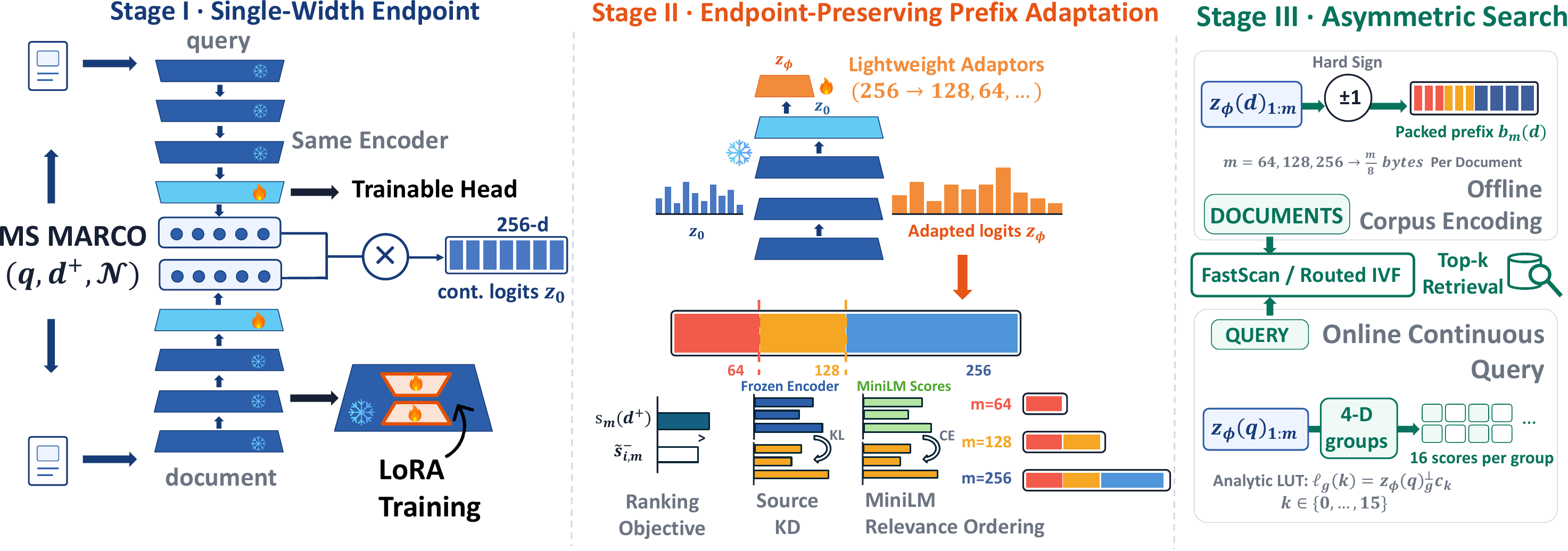}
\caption{MHR overview. Stage I trains a 256-bit source code. Stage II freezes it and trains a zero-initialized two-block residual adaptor cascade with a full-width rank anchor and 8/16/32-byte prefix losses. Deployment stores one nested hard code.}
\label{fig:overview}
\end{figure*}

\subsection{Problem and Nested Binary Codes}

Let $\mathcal{S}=\{(q_i,d_i^+,\mathcal{N}_i)\}$ be a labeled source collection and $\mathcal{D}_t$ a target corpus that is unavailable during training. We learn one source model whose document representation can be stored and searched at several byte budgets without fitting a target-dependent projection, centroid, or checkpoint. This matches deployments where the target corpus is unavailable at training time or drifts from the training distribution, the regime that BEIR-style zero-shot evaluation is designed to probe \cite{thakur2021beir}; compression also aids transfer when it can be fitted on target embeddings \cite{zuo2026compression}, which we exclude.

Let $f_{\theta_0}$ be a pretrained dual encoder. Stage I parameterizes $f_\theta$ as $\theta=\theta_0+\Delta\theta$, where only the LoRA update $\Delta\theta$ is trainable and adds a linear hash head $W\in\mathbf{R}^{L\times d}$. It produces the base logits
\begin{equation}
z_0(x)=Wf_\theta(x), \qquad h_{\beta,0}(x)=\tanh(\beta z_0(x)).
\end{equation}
We focus on the 8–32--byte severe-compression regime; $L=256$ yields a 32-byte code versus 3072 bytes for 768 float dimensions. Stage II transforms $z_0$ into adapted logits $z_\phi$. For each $m\in\mathcal{M}=\{64,128,256\}$, the stored document code is the ordered prefix
\begin{equation}
b_m(d)=\operatorname{sign}\!\left(z_\phi(d)_{1:m}\right).
\end{equation}
Prefix closure makes the operating points structurally consistent: $b_{64}(d)$ is the first half of $b_{128}(d)$ and the first quarter of $b_{256}(d)$. One forward pass, therefore, produces every budget, and an index may store only the desired prefix or retain the longest code and truncate it at query time.

\paragraph{Why direct multi-width training is limited.}

The direct construction trains $f_\theta$ and $W$ with a sum of prefix losses, $\mathcal{L}_{\mathrm{direct}}=\sum_{m\in\mathcal{M}}\alpha_m\mathcal{L}^{m}$ \cite{kusupati2022mrl}. It produces valid nested codes but underperforms a model trained only at the largest width: it improves the shortest code while weakening the 256-bit code. For a parameter tied to coordinate $j$, the update is
\begin{equation}
\nabla_{w_j}\mathcal{L}_{\mathrm{direct}}
=\sum_{m\in\mathcal{M}:j\leq m}\alpha_m\nabla_{w_j}\mathcal{L}^{m},
\label{eq:gradient}
\end{equation}
so early coordinates are optimized for every width, while later coordinates see only the wider ones. This exposure asymmetry is what Stage II targets. The effect is sharper in binary codes, where moving a coordinate across zero flips a stored bit at every width that contains it. Real-valued MRL shows an analogous width-dependent imbalance \cite{zhang2025smec}.

\subsection{Stage I: Single-Width Model Training}

We first optimize only the $L=256$ code. The source objective combines hard-document relevance ranking, frozen-BGE distillation over local and cached candidates, MiniLM ordering, and code-balance regularization, so every trainable encoder and hash-head parameter serves the full-width code that Stage II holds fixed. Following continuation-based deep hashing \cite{cao2017hashnet}, the encoder retains the bounded relaxation $h_{\beta,0}$ during optimization. In relevance and dual-encoder distillation, the document operand is $\operatorname{STE}(\operatorname{sign}(h_{\beta,0}(d)))$: the forward pass uses a hard sign and the backward pass uses the identity derivative, a straight-through estimator \cite{bengio2013estimating}. The query operand stays continuous, so Stage I already trains under the asymmetric scoring used at deployment.

The local teacher ranks each query's positive and mined negatives with the frozen BGE initialization. We retrieve the top 128 documents from 516,472 unique MS~MARCO train positives and sample four cached candidates per query, three of them from the top of that ranking.

\subsection{Stage II: Prefix Adaptation}

After Stage I, we freeze the encoder, its LoRA update, and $W$. A cascade of $R=2$ residual adaptors operates only on the fixed logits. Let $z^{(0)}(x)=z_0(x)$ and
\begin{equation}
\begin{aligned}
u^{(r)}(x)&=\operatorname{GELU}\!\left(V^{(r)}_1\operatorname{LN}(z^{(r-1)}(x))\right),\\
z^{(r)}(x)&=z^{(r-1)}(x)+V^{(r)}_2u^{(r)}(x),\\
z_\phi(x)&=z^{(R)}(x).
\end{aligned}
\label{eq:adaptor}
\end{equation}
Both the input and hidden width are 256. We initialize every $V^{(r)}_2$ to zero, making Equation~\ref{eq:adaptor} an exact identity at the start of Stage II. Thus, initial 256-bit query scores and document signs equal the Stage-I solution. The cascade redistributes capacity across coordinates without changing the encoder or replacing the source hash space. It produces one vector with nested prefixes, no per-document metadata, and no width-specific parameters.

\subsection{Source-Only Prefix Objective}

Training uses one positive and a set of mined negatives $\mathcal{N}_i$ for each source query. Let $h_{\beta,\phi}(x)=\tanh(\beta z_\phi(x))$. For prefix $m$, define the deployment-matched asymmetric training score $\bar{s}_{i,m}(d)=m^{-1}z_\phi(q_i)_{1:m}^{\top}\operatorname{STE}(\operatorname{sign}(h_{\beta,\phi}(d)_{1:m}))$. Following cross-encoder denoising in dense retrieval \cite{qu2021rocketqa}, we use MiniLM cross-encoder \cite{wang2020minilm} as the denoiser and exclude a negative when its score lies within a margin of the positive. Let $\mathcal{V}_i\subseteq\mathcal{N}_i$ contain the remaining valid negatives. We smooth the hardest-negative maximum with log-sum-exp and apply a RankNet-style softplus ranking loss \cite{burges2005learning}:
\begin{equation}
\begin{aligned}
\tilde{s}_{i,m}^{-}&=\tau_r\log\!\sum_{d\in\mathcal{V}_i}\exp\!\left(\bar{s}_{i,m}(d)/\tau_r\right),\\
\mathcal{L}_{\mathrm{rel}}^m&=\frac{1}{B}\sum_i\operatorname{softplus}\!\left((\tilde{s}_{i,m}^{-}-\bar{s}_{i,m}(d_i^+))/\tau_r\right).
\end{aligned}
\label{eq:rank}
\end{equation}

The frozen Stage-I code supplies the full-width anchor. Over candidate set $\mathcal{C}_i$, let $p_0(d\mid q_i,\mathcal{C}_i)$ be the softmax distribution induced by the full-width deployment score $L^{-1}z_0(q_i)^\top\operatorname{sign}(z_0(d))$, and let $p_{\phi,m}$ use the adapted prefix score. We distill the same source distribution into every adapted prefix \cite{hinton2015distilling}:
\begin{equation}
\mathcal{L}_{\mathrm{src}}^m=\frac{1}{B}\sum_i \operatorname{KL}\!\left[p_0(\cdot\mid q_i,\mathcal{C}_i)\,\|\,p_{\phi,m}(\cdot\mid q_i,\mathcal{C}_i)\right].
\label{eq:source-distill}
\end{equation}
The full-width ($m=L$) term penalizes drift in the 256-bit code, while the shorter terms transfer the source ranking into lower capacity. A MiniLM listwise term $\mathcal{L}_{\mathrm{ce}}^m$ complements this self-distillation with fine-grained source relevance and supplies the false-negative mask used only by Equation~\ref{eq:rank}.

The Stage-II objective is
\begin{equation}
\begin{aligned}
\mathcal{L}_{\mathrm{adapt}}={}&\frac{1}{|\mathcal{M}|}\sum_{m\in\mathcal{M}}\alpha_m\left(
\lambda_{\mathrm{rel}}\mathcal{L}_{\mathrm{rel}}^{m}
+\lambda_{\mathrm{src}}\mathcal{L}_{\mathrm{src}}^{m}
+\lambda_{\mathrm{ce}}\mathcal{L}_{\mathrm{ce}}^{m}\right)\\
&+\lambda_{\mathrm{bal}}\mathcal{L}_{\mathrm{bal}}
+\lambda_{\mathrm{gor}}\mathcal{L}_{\mathrm{gor}}.
\end{aligned}
\label{eq:objective}
\end{equation}
Here $\alpha_m$ controls the allocation of supervision across widths. Following the balanced-bit principle in learning to hash \cite{weiss2008spectral}, the balance loss penalizes a nonzero batch mean for query, positive, and negative codes, discouraging constant bits. For a code matrix $H$, define $G(H)$ as the mean squared off-diagonal cosine similarity between rows. Inspired by hyperspherical uniformity \cite{wang2020uniformity}, the prefix regularizer $\mathcal{L}_{\mathrm{gor}}=|\mathcal{M}|^{-1}\sum_m\rho_m\{G(H_q^{m})+G(H_+^{m})\}/2$ spreads examples by penalizing pairwise row similarity rather than coordinate-wise correlation.

\subsection{Asymmetric Search and FastScan}

Hard-signing a query discards information for no storage benefit, since queries are never persisted. MHR therefore uses the same raw continuous query logits during training and inference and scores them against hard document bits:
\begin{equation}
s_m(q,d)=\frac{1}{m}z_\phi(q)_{1:m}^{\top}b_m(d).
\label{eq:asym}
\end{equation}
This asymmetric score stores exactly $m/8$ code bytes per document before index metadata. Model parameters and the query lookup table do not scale with the corpus.

Equation~\ref{eq:asym} admits an analytic lookup decomposition. Partition a prefix into $G=m/4$ groups, let $c_{g}(d)\in\{-1,+1\}^4$ be the sign pattern stored for group $g$, and construct the table $\ell_g(k)=z_\phi(q)_g^\top c_k$ for each of the 16 possible patterns. Then
\begin{equation}
s_m(q,d)=\frac{1}{m}\sum_{g=1}^{G}\ell_g\!\left(c_g(d)\right).
\label{eq:lut}
\end{equation}
We install the 16 fixed sign patterns as an analytic PQ codebook and pass packed MHR bits to \texttt{IndexPQFastScan}. Scanning uses analytic lookup tables without reconstructing target vectors or fitting centroids. FAISS \cite{douze2026faiss} quantizes the query lookup table for SIMD scanning, producing a measured approximation to exact asymmetric scoring. We additionally use an eight-bit-group \texttt{IndexPQ} variant, whose 256 fixed sign patterns reproduce Equation~\ref{eq:asym} exactly, to validate packing and score orientation before benchmarking FastScan.

Index routing reuses the same payload and score. A flat index scans every code with Equation~\ref{eq:lut}. For IVF, a source-trained coarse quantizer routes target document floats once during offline list assignment, after which the floats are discarded. Each posting stores only its document ID and packed MHR code. We install the 16 fixed sign patterns for every four-bit group as a non-residual analytic codebook in \texttt{IndexIVFPQFastScan}; \texttt{search\_preassigned} receives routed cell IDs and scans their MHR payloads in FAISS C++ using continuous adapted logits. Graph pruning \cite{wang2025leann} and coarse-to-fine reranking reuse the payload the same way: the packed code selects a small candidate set that is then rescored with full-precision vectors, obtained by recomputation or from a float store.

\section{Experiments}

\subsection{Protocol}

\paragraph{Training and transfer.} We initialize Stage I from \texttt{BAAI/bge-base-en-v1.5} \cite{xiao2023cpack} and use CLS pooling. We train both stages on the MS~MARCO passage-ranking training split \cite{bajaj2016msmarco}. LoRA \cite{hu2022lora} has rank 16, scale 16, and dropout .05. The hash head uses a learning rate $2\!\times\!10^{-4}$ and the encoder adapters use $10^{-5}$. Stage I runs for 24 epochs of 600 steps. Stage II loads its final 256-bit checkpoint, freezes every Stage-I parameter, and trains two sequential 256-hidden-unit residual adaptors for eight epochs of 600 steps with a learning rate $2\!\times\!10^{-4}$. Both stages use a batch size of 64, EMA decay of 0.999, and three negatives mined by BM25 \cite{robertson2009probabilistic} and reranked by the MiniLM cross-encoder \cite{wang2020minilm} released through Sentence-Transformers \cite{reimers2019sbert} as \texttt{cross-encoder/ms-marco-MiniLM-L6-v2}. The final Stage-II BGE cascade supplies the MHR codes in all quality and systems comparisons. No target corpus is used to fit a codebook, adapt the model, mine negatives, or select a checkpoint.
Training and evaluation used one NVIDIA A100 80GB GPU; latency experiments used one CPU thread. Stage I takes about 29 GPU-hours and Stage II about 2 GPU-hours.

\paragraph{Objective details.} Stage I sets the relevance and local frozen-BGE coefficients to 3, the cached global-BGE coefficient to 1, the MiniLM coefficient to .5, and the balance coefficient to .01. Its relaxation sharpness increases linearly from 1 to 2.5 over the first 33\% of training. In the Stage-II objective of Equation~\ref{eq:objective}, we set $\lambda_{\mathrm{rel}}=\lambda_{\mathrm{src}}=3$, $\lambda_{\mathrm{ce}}=.5$, $\lambda_{\mathrm{bal}}=.01$, and $\lambda_{\mathrm{gor}}=1$. The prefix weights are $(\alpha_{64},\alpha_{128},\alpha_{256})=(1,.75,1.25)$. We use $\tau_r=.1$, source-code distillation temperature .1, MiniLM temperature 1, false-negative margin .5, fixed $\beta=2.5$, and GOR weights proportional to $1{:}.5{:}0$.


\paragraph{Datasets and metrics.} We evaluate on seven datasets distributed through BEIR \cite{thakur2021beir}: Quora, Natural Questions (NQ), HotpotQA, DBpedia-Entity, FEVER, TREC-COVID, and Climate-FEVER \cite{iyer2017quora,kwiatkowski2019natural,yang2018hotpotqa,hasibi2017dbpedia,thorne2018fever,voorhees2020treccovid,diggelmann2020climate}. They span duplicate detection, question answering, entity retrieval, and fact verification, providing heterogeneous tests of source-only transfer. NDCG@10 \cite{jarvelin2002cumulated} measures graded top-rank utility with logarithmic position discount, while qrel Recall@100 measures the fraction of judged relevant documents recovered in the first 100 ranks and, therefore, the coverage of a downstream shortlist. We also report the unweighted macro across datasets. Significance uses a paired per-query bootstrap with 10{,}000 resamples.

\paragraph{Baselines and storage.} PQ \cite{jegou2011ivfpq} and OPQ \cite{ge2013opq} use 8-bit subquantizers and codebooks fitted to MS~MARCO embeddings. Source-fitted components use 20K random MS~MARCO passages; we average five disjoint samples (Appendix~\ref{app:fitvar}). RaBitQ \cite{gao2024rabitq} uses source-fitted PCA (to 64 dimensions at 16 bytes and 192 at 32 bytes), a fixed random rotation, sign bits, and 8 persisted factor bytes per document; hence its minimum payload is 16 bytes. Our JPQ-FT reproduction \cite{zhan2021jpq} uses the same BGE initialization and MS~MARCO training data, 32 eight-bit subquantizers, 50K optimization steps, and full encoder fine-tuning. JPQ-Nested is our controlled extension that jointly optimizes the hierarchical centroid codes for the same budgets. ITQ~\cite{gong2011iterative} applies PCA, a learned Procrustes rotation, and sign binarization to frozen BGE embeddings. PCA-RR replaces the learned ITQ rotation with a fixed random orthogonal rotation after source-fitted centering and PCA, with each budget constructed independently. Signed random projection LSH~\cite{charikar2002similarity} (SRP-LSH) signs a fixed Gaussian projection, yielding a data-independent angular-LSH baseline. Super-Bit LSH~\cite{ji2012superbit} orthogonalizes the maximum-width random frame to reduce angular-estimation variance while retaining the same nested binary payloads. We report the information-theoretic payload ($m/8$ bytes) separately from serialized bytes per document, which additionally include the 64-bit document ID and any routing metadata; any full-precision vectors used by graph pruning or float reranking are accounted for separately.

\begin{figure}[t]
\centering
\includegraphics[width=0.8\columnwidth]{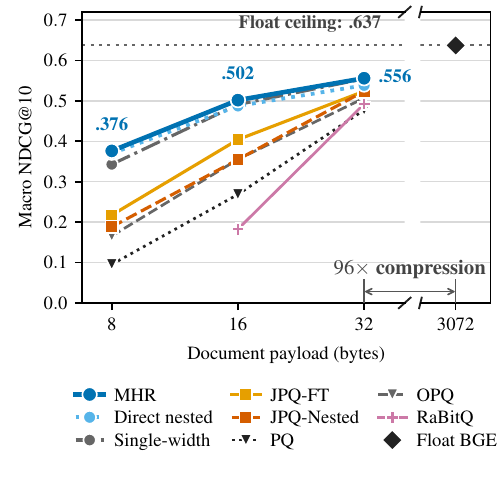}
\caption{Seven-dataset macro NDCG@10 against document payload; the axis break separates 3072-byte float reference.}
\label{fig:pareto}
\end{figure}

\subsection{Zero-Shot Quality}

Figure~\ref{fig:pareto} isolates the full-width--prefix trade-off. Single-width training gives a strong 32-byte code but weak truncated prefixes, while direct nesting improves the shortest prefix at the cost of full-width quality. MHR removes the trade-off: its 32-byte NDCG is .556, above the single-width model, and its 8- and 16-byte prefixes are the strongest of any method. Table~\ref{tab:budget} adds ITQ, PCA-RR, SRP-LSH, and Super-Bit; the figure plots the strongest external curves alongside the variants that expose the optimization effect.

Table~\ref{tab:budget} shows that the advantage holds across the storage curve. MHR has the best macro NDCG and Recall among compact baselines at every budget: at 16 bytes, it reaches .502 NDCG versus .404 for JPQ-FT. At 32 bytes, where JPQ-FT is trained directly rather than truncated, MHR improves macro NDCG / Recall from .524 / .638 to .556 / .653. All MHR points come from one prefix-closed code. Over three Stage-II seeds MHR gives .370/.499/.553 ($\pm$.006/.003/.003), and every margin over JPQ is significant ($p<.001$; Appendix~\ref{app:seeds}). Off-the-shelf binary Matryoshka embeddings trail MHR at every budget despite stronger float encoders (Appendix~\ref{app:mrl}).

\begin{table}[thbp]
\centering
{\small
\begin{tabular}{@{}lccc@{}}
\toprule
Method & 8 B & 16 B & 32 B \\
\midrule
PQ & .095 / .213 & .268 / .440 & .478 / .620 \\
OPQ & .166 / .337 & .355 / .536 & .507 / .648 \\
RaBitQ & -- & .183 / .312 & .492 / .617 \\
SRP-LSH & .097 / .169 & .212 / .316 & .366 / .491 \\
Super-Bit & .099 / .176 & .217 / .331 & .383 / .507 \\
PCA-RR & .146 / .241 & .303 / .441 & .462 / .585 \\
ITQ & .143 / .243 & .305 / .443 & .462 / .589 \\
JPQ-FT & .217 / .342 & .404 / .537 & .524 / .638 \\
JPQ-Nested & .188 / .324 & .355 / .516 & .524 / .643 \\
\textbf{MHR} & \textbf{.376 / .518} & \textbf{.502 / .616} & \textbf{.556 / .653} \\
\midrule
Float BGE & \multicolumn{3}{c}{.637 / .720} \\
\bottomrule
\end{tabular}}
\caption{Seven-dataset macro NDCG@10 / Recall@100. Source-fitted rows average five MS~MARCO fitting samples (std $\le$.01). JPQ-FT at 8/16\,B truncates its 32\,B model.}
\label{tab:budget}
\end{table}

\begin{table*}[thbp]
\centering
\small
\renewcommand{\arraystretch}{.92}
\begin{minipage}[t]{.485\textwidth}
\centering
{\small
\textit{(a) 32 B, NDCG@10}\par\vspace{2pt}
\setlength{\tabcolsep}{4pt}
\begin{tabular}{@{}lrrrrrrrr@{}}
\toprule
\textbf{Method} & \textbf{Q} & \textbf{NQ} & \textbf{HP} & \textbf{DB} & \textbf{FV} & \textbf{TC} & \textbf{CF} & \textbf{Avg.} \\
\midrule
Float & .888 & .536 & .710 & .405 & .852 & .780 & .289 & .637 \\
\midrule
PQ & .828 & .363 & .468 & .295 & .638 & .559 & .190 & .478 \\
OPQ & .832 & .402 & .497 & .328 & \underline{.680} & .602 & \underline{.206} & .507 \\
LSH & .814 & .267 & .299 & .173 & .478 & .450 & .084 & .366 \\
S-Bit & .819 & .281 & .327 & .188 & .515 & .466 & .086 & .383 \\
PCA-RR & \underline{.840} & .355 & .395 & .264 & .610 & .604 & .166 & .462 \\
ITQ & \underline{.840} & .362 & .390 & .261 & .605 & .612 & .168 & .462 \\
JPQ-FT & .839 & \underline{.433} & \textbf{.545} & .338 & .661 & \underline{.686} & .166 & \underline{.524} \\
JPQ-N & .835 & \textbf{.436} & .538 & \underline{.344} & .662 & .678 & .174 & \underline{.524} \\
\textbf{MHR} & \textbf{.850} & .431 & \underline{.544} & \textbf{.355} & \textbf{.762} & \textbf{.704} & \textbf{.246} & \textbf{.556} \\
\bottomrule
\end{tabular}}
\end{minipage}\hfill
\begin{minipage}[t]{.485\textwidth}
\centering
{\small
\textit{(b) 32 B, Recall@100}\par\vspace{2pt}
\setlength{\tabcolsep}{4pt}
\begin{tabular}{@{}lrrrrrrrr@{}}
\toprule
\textbf{Method} & \textbf{Q} & \textbf{NQ} & \textbf{HP} & \textbf{DB} & \textbf{FV} & \textbf{TC} & \textbf{CF} & \textbf{Avg.} \\
\midrule
Float & .997 & .939 & .852 & .524 & .969 & .141 & .616 & .720 \\
\midrule
PQ & \underline{.990} & .792 & .694 & .369 & .911 & .086 & .497 & .620 \\
OPQ & \textbf{.992} & .844 & \textbf{.729} & .424 & \underline{.931} & .102 & \underline{.517} & \underline{.648} \\
LSH & .979 & .642 & .505 & .221 & .788 & .060 & .237 & .491 \\
S-Bit & .981 & .673 & .541 & .249 & .816 & .065 & .222 & .507 \\
PCA-RR & .989 & .791 & .615 & .332 & .885 & .096 & .389 & .585 \\
ITQ & .989 & .794 & .614 & .337 & .886 & .100 & .402 & .589 \\
JPQ-FT & \textbf{.992} & \underline{.883} & \underline{.704} & \underline{.434} & .929 & \underline{.119} & .405 & .638 \\
JPQ-N & \underline{.990} & \textbf{.885} & \underline{.704} & \textbf{.437} & .928 & \underline{.119} & .435 & .643 \\
\textbf{MHR} & \underline{.990} & .868 & .695 & .427 & \textbf{.940} & \textbf{.121} & \textbf{.532} & \textbf{.653} \\
\bottomrule
\end{tabular}}
\end{minipage}

\vspace{2pt}

\begin{minipage}[t]{.485\textwidth}
\centering
{\small
\textit{(c) 16 B, NDCG@10}\par\vspace{2pt}
\setlength{\tabcolsep}{4pt}
\begin{tabular}{@{}lrrrrrrrr@{}}
\toprule
\textbf{Method} & \textbf{Q} & \textbf{NQ} & \textbf{HP} & \textbf{DB} & \textbf{FV} & \textbf{TC} & \textbf{CF} & \textbf{Avg.} \\
\midrule
Float & .888 & .536 & .710 & .405 & .852 & .780 & .289 & .637 \\
\midrule
PQ & .703 & .145 & .157 & .144 & .268 & .349 & .114 & .268 \\
OPQ & .757 & .253 & .245 & .215 & .424 & .448 & \underline{.146} & .355 \\
LSH & .703 & .123 & .103 & .068 & .184 & .276 & .025 & .212 \\
S-Bit & .714 & .133 & .121 & .070 & .203 & .256 & .021 & .217 \\
PCA-RR & .773 & .207 & .153 & .143 & .308 & .450 & .085 & .303 \\
ITQ & .772 & .203 & .147 & .149 & .312 & .458 & .098 & .305 \\
JPQ-FT & \underline{.788} & \underline{.329} & \underline{.360} & \underline{.259} & \underline{.480} & \underline{.502} & .111 & \underline{.404} \\
JPQ-N & .740 & .297 & .270 & .225 & .379 & .456 & .118 & .355 \\
\textbf{MHR} & \textbf{.827} & \textbf{.377} & \textbf{.449} & \textbf{.315} & \textbf{.690} & \textbf{.642} & \textbf{.214} & \textbf{.502} \\
\bottomrule
\end{tabular}}
\end{minipage}\hfill
\begin{minipage}[t]{.485\textwidth}
\centering
{\small
\textit{(d) 16 B, Recall@100}\par\vspace{2pt}
\setlength{\tabcolsep}{4pt}
\begin{tabular}{@{}lrrrrrrrr@{}}
\toprule
\textbf{Method} & \textbf{Q} & \textbf{NQ} & \textbf{HP} & \textbf{DB} & \textbf{FV} & \textbf{TC} & \textbf{CF} & \textbf{Avg.} \\
\midrule
Float & .997 & .939 & .852 & .524 & .969 & .141 & .616 & .720 \\
\midrule
PQ & .959 & .473 & .374 & .203 & .630 & .056 & .386 & .440 \\
OPQ & .978 & .676 & .508 & .288 & .783 & .075 & \underline{.443} & .536 \\
LSH & .928 & .381 & .245 & .093 & .449 & .031 & .085 & .316 \\
S-Bit & .936 & .405 & .270 & .103 & .478 & .033 & .089 & .331 \\
PCA-RR & .971 & .593 & .351 & .204 & .636 & .070 & .258 & .441 \\
ITQ & .970 & .586 & .346 & .204 & .642 & .075 & .280 & .443 \\
JPQ-FT & \underline{.981} & \underline{.738} & \underline{.537} & \underline{.305} & \underline{.815} & \underline{.078} & .306 & \underline{.537} \\
JPQ-N & .976 & .709 & .474 & .287 & .757 & .073 & .339 & .517 \\
\textbf{MHR} & \textbf{.986} & \textbf{.821} & \textbf{.617} & \textbf{.379} & \textbf{.921} & \textbf{.108} & \textbf{.481} & \textbf{.616} \\
\bottomrule
\end{tabular}}
\end{minipage}
\caption{Per-dataset quality at 16 and 32 bytes; bold/underline: best/second-best compact method. Q: Quora, HP: HotpotQA, DB: DBpedia-Entity, FV: FEVER, TC: TREC-COVID, CF: Climate-FEVER; JPQ-N: JPQ-Nested; S-Bit: Super-Bit.}
\label{tab:dataset}
\end{table*}

Table~\ref{tab:dataset} shows that the macro gains are not driven by one corpus (13 datasets and in-domain results: Appendix~\ref{app:datasets}). At 16 bytes, MHR leads every compact baseline on all seven datasets in both metrics. At 32 bytes, it is best on five datasets in NDCG, with the largest gains on the fact-verification tasks and TREC-COVID. MHR keeps 87\% of float BGE's macro NDCG at a 96$\times$ smaller payload. The gains also hold on another backbone: MHR trained on nomic-embed-v1.5 beats its own binary MRL by .14/.11/.05 at 8/16/32 bytes (Appendix~\ref{app:mrl}).

\subsection{Ablation by Component}
\label{subsec:component}

Table~\ref{tab:ablations} removes one ingredient at a time. Without $\mathcal{L}_{\mathrm{src}}$, mean NDCG drops by .133 and mean Recall by .125, below the direct-nested and single-width-truncated references (.488/.484): without the full-width anchor, the adaptor drifts off the source solution. $\mathcal{L}_{\mathrm{rel}}$ contributes .013. A second block and raw-logit training each add .007, neither reaches .498 alone, and a third block does not help. $\mathcal{L}_{\mathrm{ce}}$, $\mathcal{L}_{\mathrm{gor}}$, and tuned $\alpha$ each add .002--.004. Removing the bit-balance or GOR term significantly lowers 16-byte NDCG, and no other setting beats the default (Appendix~\ref{app:sens}); Appendix~\ref{app:mrlctrl} tests an MRL-style Stage I.

Distilling every width reaches .472 mean NDCG, versus .466 for the full-width term alone, .458 for the two shorter widths alone, and .438 for no distillation. An inference-only control on the direct-nested checkpoint tests the query representation from the other side: replacing raw query logits with hard query signs lowers macro NDCG by .033 to .088 across widths, most at 8 bytes. Continuous query scoring recovers this quality at no storage cost.

\begin{table*}[t]
\centering
\small
\begin{minipage}[t]{.49\textwidth}
\centering
{\small
\textit{(a) Adaptor design}\par\vspace{2pt}
\setlength{\tabcolsep}{2.5pt}
\begin{tabular}{@{}lcccc@{}}
\toprule
Variant & 8 B & 16 B & 32 B & Mean \\
\midrule
\textbf{MHR (selected)} & \textbf{.396/.489} & \textbf{.522/.582} & \textbf{.577/.620} & \textbf{.498} \\
one block & .387/.487 & .516/.584 & .569/.622 & .491 \\
tanh query & .388/.486 & .516/.580 & .568/.620 & .491 \\
uniform $\alpha$ & .386/.488 & .519/.583 & .577/.621 & .494 \\
\bottomrule
\end{tabular}}
\end{minipage}\hfill
\begin{minipage}[t]{.49\textwidth}
\centering
{\small
\textit{(b) Objective terms}\par\vspace{2pt}
\setlength{\tabcolsep}{2.5pt}
\begin{tabular}{@{}lcccc@{}}
\toprule
Removed & 8 B & 16 B & 32 B & Mean \\
\midrule
$\mathcal{L}_{\mathrm{src}}$ & .295/.389 & .379/.449 & .423/.480 & .365 \\
$\mathcal{L}_{\mathrm{rel}}$ & .380/.476 & .507/.575 & .567/.619 & .485 \\
$\mathcal{L}_{\mathrm{ce}}$ & .393/.486 & .522/.581 & .573/.622 & .496 \\
$\mathcal{L}_{\mathrm{gor}}$ & .389/.489 & .520/.582 & .576/.620 & .495 \\
\bottomrule
\end{tabular}}
\end{minipage}
\caption{Stage-II ablations on Quora, NQ, HotpotQA, DBpedia-Entity, and TREC-COVID: NDCG@10/Recall@100, with Mean averaging NDCG over widths. (a) changes one design choice; (b) removes one loss term.}
\label{tab:ablations}
\end{table*}

\begin{table*}[t]
\centering
\begin{minipage}[t]{.49\textwidth}
\centering
{\small
\textit{(a) Flat scan}\par\vspace{2pt}
\setlength{\tabcolsep}{3.0pt}
\begin{tabular}{@{}lrrrrrr@{}}
\toprule
& \multicolumn{3}{c}{Quora} & \multicolumn{3}{c}{TREC-C} \\
Method & B/doc & NDCG & p50 & B/doc & NDCG & p50 \\
\midrule
MHR FastScan & 32.0 & .8501 & .490 & 32.1 & .7075 & .192 \\
\quad prefix cascade & 32.0 & .8495 & \textbf{.132} & 32.0 & .7025 & \textbf{.118} \\
OPQ FastScan & 36.6 & .8340 & .502 & 46.1 & .5791 & .205 \\
PQ flat & 33.5 & .8284 & 4.413 & 36.6 & .5593 & 1.467 \\
Float flat & 3072 & .8876 & 7.758 & 3072 & .7802 & 3.511 \\
\bottomrule
\end{tabular}}
\end{minipage}\hfill
\begin{minipage}[t]{.49\textwidth}
\centering
{\small
\textit{(b) Routed IVF scan}\par\vspace{2pt}
\setlength{\tabcolsep}{2.2pt}
\begin{tabular}{@{}c*{8}{r}@{}}
\toprule
& \multicolumn{4}{c}{Quora}
& \multicolumn{4}{c}{TREC-C} \\
\cmidrule(lr){2-5}\cmidrule(lr){6-9}
& \multicolumn{2}{c}{NDCG}
& \multicolumn{2}{c}{ms/q}
& \multicolumn{2}{c}{NDCG}
& \multicolumn{2}{c}{ms/q} \\
$n_{\rm probe}$
& MHR & PQ & MHR & PQ
& MHR & PQ & MHR & PQ \\
\midrule
1
& \textbf{.603} & .588 & \textbf{.021} & .022
& \textbf{.596} & .483 & \textbf{.031} & .032 \\
4
& \textbf{.800} & .762 & \textbf{.040} & .042
& \textbf{.685} & .538 & \textbf{.061} & .063 \\
16
& \textbf{.844} & .792 & \textbf{.088} & .090
& \textbf{.709} & .543 & \textbf{.127} & .133 \\
64
& \textbf{.850} & .794 & \textbf{.231} & .235
& \textbf{.707} & .543 & \textbf{.191} & .198 \\
\bottomrule
\end{tabular}}
\end{minipage}
\caption{Search efficiency, one CPU thread; latency is the median per-query time over five trials. (a) Flat scan; B/doc is serialized storage. (b) IVF with FastScan postings for both methods; latency includes routing. Source-fitted cells average five samples.}
\label{tab:systems}
\end{table*}

\subsection{Scan Efficiency}

Table~\ref{tab:systems}(a) shows single-thread MHR FastScan at .49/.19 ms per query on Quora/TREC-COVID, 9.0$\times$/7.7$\times$ faster than source-fitted PQ flat. Scoring the same document bits by exact Hamming distance is both less accurate (Section~\ref{subsec:component}) and 1.57$\times$/1.50$\times$ slower. A per-document XOR–popcount is cheaper than a table lookup, but the asymmetric score maps onto FAISS \texttt{IndexPQFastScan}, a register-blocked SIMD lookup kernel that scans 32 codes at once, whereas exact Hamming uses the less vectorized popcount loop of \texttt{IndexBinaryFlat}. At matched FastScan kernel and latency, MHR is more accurate than source-fitted OPQ. Nesting adds a speedup that PQ codes cannot provide: a prefix cascade (a C kernel; other rows use FAISS) scans the 16-byte prefix and re-scores a shortlist with the full code, 3.7$\times$/1.6$\times$ faster within .005 NDCG (Appendix~\ref{app:cascade}).

We next integrate MHR in IVF~\cite{jegou2011ivfpq}, a routing-based index, to replace PQ. IVF-MHR carries the MHR code as the posting-list payload of a FAISS \texttt{IndexIVFPQFastScan}, scored in-list with the analytic asymmetric lookup of Equation~\ref{eq:asym}; IVF-PQ is the standard FAISS index with the same FastScan layout and a PQ codebook fitted on source residuals. Both use the same MS~MARCO-trained coarse quantizer ($n_\mathrm{list}=256$) and a 32-byte payload. With 64-bit IDs and routing metadata, serialized MHR/PQ storage is 42.3/41.8 B/doc on Quora and 47.1/45.7 B/doc on TREC-COVID, so the serialized MHR index is 72.6$\times$/65.2$\times$ smaller than float vectors (the 96$\times$ ratio counts the code payload alone).

Table~\ref{tab:systems}(b) reports quality and end-to-end latency at four probe counts. PQ saturates as $n_\mathrm{probe}$ grows: from 4 to 64 probes, its NDCG gains .032 on Quora and .005 on TREC-COVID, against .050 and .022 for MHR. With identical routing and payload, MHR ranks documents within a list better: at one probe on TREC-COVID it already exceeds PQ's best result (.596 vs.\ .543). At all eight operating points, IVF-MHR has higher NDCG than IVF-PQ at near-equal serialized storage and is up to 4\% faster. At 16 probes (.127/.088 ms), IVF-MHR matches flat-scan NDCG on TREC-COVID and is within .006 on Quora.

\subsection{Graph-Guided Selective Recomputation}

LEANN keeps a compact code at each HNSW node to prune neighbor expansions and recomputes full embeddings only for selected nodes \cite{wang2025leann}. In our 8-byte LEANN configuration, we replace its PQ code with the 8-byte raw-query MHR prefix. All methods share the graph ($M=32$, $\mathrm{efConstruction}=200$), BGE exact distances, $\mathrm{efSearch}=128$, and LEANN's global pruning.

Unpruned search makes 2206/2624/3521 exact distance evaluations per query on TREC-COVID/Quora/NQ. At 25\% retention, MHR keeps 99.6--100.0\% of unpruned NDCG@10 and 97.4--99.8\% of Recall@100 with 66.6--68.6\% fewer evaluations (Table~\ref{tab:leann}).
It outperforms both PQ controls on Quora and NQ, by wide margins on NQ (.529 vs .444 NDCG against \textit{target-fitted} PQ), and ties \textit{target-fitted} PQ on TREC-COVID despite \textit{never} seeing the target corpus. On Quora and NQ it also needs fewer exact evaluations than either PQ code. Because LEANN recomputes each embedding for lower storage overhead, the search cost is dominated by this evaluation count. MHR ranks neighbors almost as well as unpruned search, so LEANN can prune aggressively without losing quality.

\begin{table}[thbp]
\centering
{\small
\setlength{\tabcolsep}{1.0pt}
\begin{tabular}{@{}lccccr@{}}
\toprule
& \multicolumn{4}{c}{NDCG@10 / Recall@100} & \\
Data & Unpr. & PQ-T & PQ-S & MHR & Exact \\
\midrule
TREC-C & .780/.142 & \textbf{.779}/.129 & .773/.128 & .777/\textbf{.139} & 689/776/737 \\
Quora & .887/.996 & .867/.962 & .883/.985 & \textbf{.887/.994} & 837/852/834 \\
NQ & .531/.929 & .444/.702 & .452/.716 & \textbf{.529/.909} & 1180/1200/1105 \\
\bottomrule
\end{tabular}}
\caption{LEANN-HNSW pruning with 8-byte codes at 25\% retention. Exact: full-vector distance evaluations per query (PQ-T/PQ-S/MHR). PQ-T is target-fitted; PQ-S is source-fitted (mean of five samples).}
\label{tab:leann}
\end{table}

\subsection{Compact Coarse Filtering with Reranking}

A nested MHR prefix can also serve as a first-stage filter: scan all documents at a narrow width, keep the top $K$, and rerank them with full-precision embeddings (Table~\ref{tab:coarse}). At $K=100$, 8-byte MHR reaches .561 versus .185 for PQ, exceeding its 32-byte compact score (.552); at 32 bytes, it reaches .629, near the .637 float ceiling. The adaptor improves the 32-byte $K=100$ macro over direct nesting from .6240 to .6289, while both converge as $K$ grows.

\begin{table}[thbp]
\centering
{\small
\setlength{\tabcolsep}{3.0pt}
\begin{tabular}{@{}lcccc@{}}
\toprule
Method & $K=100$ & $K=500$ & $K=1000$ & Compact only \\
\midrule
MHR 8B  & .561 & .596 & .608 & .363 \\
PQ 8B   & .185 & .303 & .354 & .045 \\
MHR 16B & .616 & .628 & .630 & .500 \\
PQ 16B  & .461 & .546 & .577 & .217 \\
MHR 32B & .629 & .633 & .634 & .552 \\
PQ 32B  & .606 & .628 & .632 & .429 \\
\midrule
Float BGE & \multicolumn{4}{c}{.637} \\
\bottomrule
\end{tabular}}
\caption{Coarse filtering with $K$ float-reranked candidates: seven-dataset macro NDCG@10.}
\label{tab:coarse}
\end{table}

\section{Concluding Remarks}

MHR replaces target codebook fitting and corpus re-encoding with one source-trained nested code. At 32 bytes, it improves macro NDCG by .032 and Recall by .016 over JPQ-FT at a 96$\times$ payload reduction. A two-block residual adaptor cascade over the frozen source code preserves the 256-bit code while organizing one 8/16/32-byte prefix. Across seven zero-shot datasets, MHR outperforms compact baselines at every budget; FastScan and IVF keep these gains at compact-index latency, a prefix cascade over the same stored code cuts flat-scan latency by up to 3.7$\times$, 8-byte codes improve LEANN-HNSW graph pruning over source-fitted PQ while staying near the unpruned ceiling, and an 8-byte coarse filter recovers near-float quality after reranking a few candidates.

This study focuses on in-memory CPU search. Extending MHR to FAISS-style GPU flat and IVF kernels \cite{johnson2021gpu} and disk-resident graph search, such as DiskANN \cite{subramanya2019diskann}, is a natural next step.

Recent private retrieval systems also build on compact binary codes: to shortlist candidates before protected reranking~\cite{hua2026pointing}, to reduce secure-computation cost through compact scoring and filtering~\cite{hua2026spruce}, or to search perturbed codes without heavy cryptography~\cite{cui2026mess}. Nested MHR prefixes would let them trade quality, privacy, and cost within one code.

\bibliography{references}

@inproceedings{karpukhin2020dpr,
  author = {Vladimir Karpukhin and Barlas Oguz and Sewon Min and Patrick Lewis and Ledell Wu and Sergey Edunov and Danqi Chen and Yih, Wen-tau},
  title = {Dense Passage Retrieval for Open-Domain Question Answering},
  booktitle = {Proceedings of the 2020 Conference on Empirical Methods in Natural Language Processing},
  pages = {6769--6781},
  year = {2020},
  doi = {10.18653/v1/2020.emnlp-main.550}
}

@inproceedings{hua2026semantics,
  title={Semantics Delivery Network: Rethinking Web Retrieval Infrastructure for LLM Agents},
  author={Peichun Hua and Yunming Xiao},
  booktitle={The 25th ACM Workshop on Hot Topics in Networks (HotNets '26)},
  year={2026},
  publisher={Association for Computing Machinery},
  address={New York, NY, USA},
  url={https://doi.org/10.1145/3847290.3859940},
  doi={10.1145/3847290.3859940},
  numpages={13},
  location={Salt Lake City, UT, USA},
  series={HotNets '26}
}

@inproceedings{lewis2020rag,
  author = {Patrick Lewis and Ethan Perez and Aleksandra Piktus and Fabio Petroni and Vladimir Karpukhin and Naman Goyal and Heinrich K{\"u}ttler and Mike Lewis and Wen{-}tau Yih and Tim Rockt{\"a}schel and Sebastian Riedel and Douwe Kiela},
  title = {Retrieval-Augmented Generation for Knowledge-Intensive {NLP} Tasks},
  booktitle = {Advances in Neural Information Processing Systems},
  volume = {33},
  year = {2020}
}

@inproceedings{xiao2023cpack,
  author = {Shitao Xiao and Zheng Liu and Peitian Zhang and Niklas Muennighoff and Defu Lian and Jian{-}Yun Nie},
  title = {{C-Pack}: Packed Resources for General Chinese Embeddings},
  booktitle = {Proceedings of the 47th International ACM SIGIR Conference on Research and Development in Information Retrieval},
  pages = {641--649},
  publisher = {ACM},
  year = {2024},
  doi = {10.1145/3626772.3657878}
}

@article{bajaj2016msmarco,
  author = {Payal Bajaj and Daniel Campos and Nick Craswell and Li Deng and Jianfeng Gao and Xiaodong Liu and Rangan Majumder and Andrew McNamara and Bhaskar Mitra and Tri Nguyen and Mir Rosenberg and Xia Song and Alina Stoica and Saurabh Tiwary and Tong Wang},
  title = {{MS MARCO}: A Human Generated Machine Reading Comprehension Dataset},
  journal = {CoRR},
  volume = {abs/1611.09268},
  year = {2016},
  doi = {10.48550/arXiv.1611.09268}
}

@inproceedings{thakur2021beir,
  author = {Nandan Thakur and Nils Reimers and Andreas R{\"u}ckl{\'e} and Abhishek Srivastava and Iryna Gurevych},
  title = {{BEIR}: A Heterogeneous Benchmark for Zero-shot Evaluation of Information Retrieval Models},
  booktitle = {Proceedings of the Neural Information Processing Systems Track on Datasets and Benchmarks},
  volume = {1},
  year = {2021}
}

@inproceedings{thakur2023domainhash,
  author = {Nandan Thakur and Nils Reimers and Jimmy Lin},
  title = {Injecting Domain Adaptation with Learning-to-Hash for Effective and Efficient Zero-Shot Dense Retrieval},
  booktitle = {Proceedings of the Workshop on Reaching Efficiency in Neural Information Retrieval},
  year = {2023},
  url = {https://arxiv.org/abs/2205.11498},
  note = {ReNeuIR at SIGIR 2023}
}

@article{jegou2011ivfpq,
  author = {Herv{\'e} J{\'e}gou and Matthijs Douze and Cordelia Schmid},
  title = {Product Quantization for Nearest Neighbor Search},
  journal = {IEEE Transactions on Pattern Analysis and Machine Intelligence},
  volume = {33},
  number = {1},
  pages = {117--128},
  year = {2011},
  doi = {10.1109/TPAMI.2010.57}
}

@inproceedings{ge2013opq,
  author = {Tiezheng Ge and Kaiming He and Qifa Ke and Jian Sun},
  title = {Optimized Product Quantization for Approximate Nearest Neighbor Search},
  booktitle = {Proceedings of the IEEE Conference on Computer Vision and Pattern Recognition},
  pages = {2946--2953},
  year = {2013},
  doi = {10.1109/CVPR.2013.379}
}

@inproceedings{gong2011iterative,
  author = {Yunchao Gong and Svetlana Lazebnik},
  title = {Iterative Quantization: A Procrustean Approach to Learning Binary Codes},
  booktitle = {Proceedings of the IEEE Conference on Computer Vision and Pattern Recognition},
  pages = {817--824},
  year = {2011},
  doi = {10.1109/CVPR.2011.5995432}
}

@inproceedings{charikar2002similarity,
  author = {Moses Charikar},
  title = {Similarity Estimation Techniques from Rounding Algorithms},
  booktitle = {Proceedings of the Thirty-Fourth Annual ACM Symposium on Theory of Computing},
  pages = {380--388},
  year = {2002},
  doi = {10.1145/509907.509965}
}

@article{malkov2018hnsw,
  author = {Yu. A. Malkov and D. A. Yashunin},
  title = {Efficient and Robust Approximate Nearest Neighbor Search Using Hierarchical Navigable Small World Graphs},
  journal = {IEEE Transactions on Pattern Analysis and Machine Intelligence},
  volume = {42},
  number = {4},
  pages = {824--836},
  year = {2020},
  doi = {10.1109/TPAMI.2018.2889473}
}

@inproceedings{subramanya2019diskann,
  author = {Suhas Jayaram Subramanya and Fnu Devvrit and Harsha Vardhan Simhadri and Ravishankar Krishnawamy and Rohan Kadekodi},
  title = {{DiskANN}: Fast Accurate Billion-point Nearest Neighbor Search on a Single Node},
  booktitle = {Advances in Neural Information Processing Systems},
  volume = {32},
  year = {2019}
}

@article{wang2025leann,
  author = {Yichuan Wang and Zhifei Li and Shu Liu and Yongji Wu and Ziming Mao and Yilong Zhao and Xiao Yan and Zhiying Xu and Yang Zhou and Ion Stoica and Sewon Min and Matei Zaharia and Joseph E. Gonzalez},
  title = {{LEANN}: A Low-Storage Vector Index},
  journal = {CoRR},
  volume = {abs/2506.08276},
  year = {2025},
  doi = {10.48550/arXiv.2506.08276}
}

@inproceedings{kusupati2022mrl,
  author = {Aditya Kusupati and Gantavya Bhatt and Aniket Rege and Matthew Wallingford and Aditya Sinha and Vivek Ramanujan and William Howard-Snyder and Kaifeng Chen and Sham M. Kakade and Prateek Jain and Ali Farhadi},
  title = {Matryoshka Representation Learning},
  booktitle = {Advances in Neural Information Processing Systems},
  volume = {35},
  year = {2022}
}

@inproceedings{yoon2024matryoshka,
  author = {Jinsung Yoon and Rajarishi Sinha and Sercan O. Arik and Tomas Pfister},
  title = {Matryoshka-Adaptor: Unsupervised and Supervised Tuning for Smaller Embedding Dimensions},
  booktitle = {Proceedings of the 2024 Conference on Empirical Methods in Natural Language Processing},
  pages = {10318--10336},
  publisher = {Association for Computational Linguistics},
  year = {2024},
  doi = {10.18653/v1/2024.emnlp-main.576}
}

@inproceedings{zhang2025smec,
  author = {Biao Zhang and Lixin Chen and Tong Liu and Bo Zheng},
  title = {{SMEC}: Rethinking Matryoshka Representation Learning for Retrieval Embedding Compression},
  booktitle = {Proceedings of the 2025 Conference on Empirical Methods in Natural Language Processing},
  pages = {26209--26222},
  publisher = {Association for Computational Linguistics},
  year = {2025},
  doi = {10.18653/v1/2025.emnlp-main.1332}
}

@inproceedings{liu2019jmh,
  author = {Xingbo Liu and Xiushan Nie and Yingxin Wang and Yilong Yin},
  title = {Jointly Multiple Hash Learning},
  booktitle = {Proceedings of the Thirty-Third AAAI Conference on Artificial Intelligence},
  volume = {33},
  pages = {9981--9982},
  year = {2019},
  doi = {10.1609/aaai.v33i01.33019981}
}

@article{luo2021mah,
  author = {Yadan Luo and Zi Huang and Yang Li and Fumin Shen and Yang Yang and Peng Cui},
  title = {Collaborative Learning for Extremely Low Bit Asymmetric Hashing},
  journal = {IEEE Transactions on Knowledge and Data Engineering},
  volume = {33},
  number = {12},
  pages = {3675--3685},
  year = {2021},
  doi = {10.1109/TKDE.2020.2977633}
}

@article{he2024nhl,
  author = {Liyang He and Yuren Zhang and Rui Li and Zhenya Huang and Runze Wu and Enhong Chen},
  title = {Nested Hash Layer: A Plug-and-play Module for Multiple-length Hash Code Learning},
  journal = {CoRR},
  volume = {abs/2412.08922},
  year = {2024},
  doi = {10.48550/arXiv.2412.08922}
}

@inproceedings{rege2023adanns,
  author = {Aniket Rege and Aditya Kusupati and Sharan Ranjit S and Alan Fan and Qingqing Cao and Sham Kakade and Prateek Jain and Ali Farhadi},
  title = {{AdANNS}: A Framework for Adaptive Semantic Search},
  booktitle = {Advances in Neural Information Processing Systems},
  volume = {36},
  year = {2023}
}

@article{gao2024rabitq,
  author = {Jianyang Gao and Cheng Long},
  title = {{RaBitQ}: Quantizing High-Dimensional Vectors with a Theoretical Error Bound for Approximate Nearest Neighbor Search},
  journal = {Proceedings of the ACM on Management of Data},
  volume = {2},
  number = {3},
  pages = {1--27},
  year = {2024},
  doi = {10.1145/3654970}
}

@inproceedings{zhang2026ike,
  author = {Zhibo Zhang and Yang Xu and Kai Ming Ting and Cam-Tu Nguyen},
  title = {{LLM}s Meet Isolation Kernel: Lightweight, Learning-free Binary Embeddings for Fast Retrieval},
  booktitle = {Findings of the Association for Computational Linguistics: ACL 2026},
  pages = {13601--13623},
  publisher = {Association for Computational Linguistics},
  year = {2026},
  doi = {10.18653/v1/2026.findings-acl.666}
}

@article{douze2026faiss,
  author = {Matthijs Douze and Alexandr Guzhva and Chengqi Deng and Jeff Johnson and Gergely Szilvasy and Pierre-Emmanuel Mazar{\'e} and Maria Lomeli and Lucas Hosseini and Herv{\'e} J{\'e}gou},
  title = {The {Faiss} Library},
  journal = {IEEE Transactions on Big Data},
  volume = {12},
  number = {2},
  pages = {346--361},
  year = {2026},
  doi = {10.1109/TBDATA.2025.3618474}
}

@article{johnson2021gpu,
  author = {Jeff Johnson and Matthijs Douze and Herv{\'e} J{\'e}gou},
  title = {Billion-Scale Similarity Search with {GPU}s},
  journal = {IEEE Transactions on Big Data},
  volume = {7},
  number = {3},
  pages = {535--547},
  year = {2021},
  doi = {10.1109/TBDATA.2019.2921572}
}

@inproceedings{yamada2021bpr,
  author = {Ikuya Yamada and Akari Asai and Hannaneh Hajishirzi},
  title = {Efficient Passage Retrieval with Hashing for Open-domain Question Answering},
  booktitle = {Proceedings of the 59th Annual Meeting of the Association for Computational Linguistics and the 11th International Joint Conference on Natural Language Processing},
  pages = {979--986},
  year = {2021},
  doi = {10.18653/v1/2021.acl-short.123}
}

@inproceedings{gan2023bebr,
  author = {Yukang Gan and Yixiao Ge and Chang Zhou and Shupeng Su and Zhouchuan Xu and Xuyuan Xu and Quanchao Hui and Xiang Chen and Yexin Wang and Ying Shan},
  title = {Binary Embedding-based Retrieval at Tencent},
  booktitle = {Proceedings of the 29th ACM SIGKDD Conference on Knowledge Discovery and Data Mining},
  pages = {4056--4067},
  publisher = {Association for Computing Machinery},
  year = {2023},
  doi = {10.1145/3580305.3599782}
}

@inproceedings{ou2021dhim,
  author = {Zijing Ou and Qinliang Su and Jianxing Yu and Ruihui Zhao and Yefeng Zheng and Bang Liu},
  title = {Refining {BERT} Embeddings for Document Hashing via Mutual Information Maximization},
  booktitle = {Findings of the Association for Computational Linguistics: EMNLP 2021},
  pages = {2360--2369},
  year = {2021},
  doi = {10.18653/v1/2021.findings-emnlp.203}
}

@inproceedings{zhan2021jpq,
  author = {Jingtao Zhan and Jiaxin Mao and Yiqun Liu and Jiafeng Guo and Min Zhang and Shaoping Ma},
  title = {Jointly Optimizing Query Encoder and Product Quantization to Improve Retrieval Performance},
  booktitle = {Proceedings of the 30th ACM International Conference on Information and Knowledge Management},
  pages = {2487--2496},
  publisher = {Association for Computing Machinery},
  year = {2021},
  doi = {10.1145/3459637.3482358}
}

@inproceedings{zhan2022repconc,
  author = {Jingtao Zhan and Jiaxin Mao and Yiqun Liu and Jiafeng Guo and Min Zhang and Shaoping Ma},
  title = {Learning Discrete Representations via Constrained Clustering for Effective and Efficient Dense Retrieval},
  booktitle = {Proceedings of the Fifteenth ACM International Conference on Web Search and Data Mining},
  pages = {1328--1336},
  publisher = {Association for Computing Machinery},
  year = {2022},
  doi = {10.1145/3488560.3498443}
}

@inproceedings{hu2022lora,
  author = {Edward J. Hu and Yelong Shen and Phillip Wallis and Zeyuan Allen-Zhu and Yuanzhi Li and Shean Wang and Lu Wang and Weizhu Chen},
  title = {{LoRA}: Low-Rank Adaptation of Large Language Models},
  booktitle = {International Conference on Learning Representations},
  year = {2022}
}

@inproceedings{wang2020minilm,
  author = {Wenhui Wang and Furu Wei and Li Dong and Hangbo Bao and Nan Yang and Ming Zhou},
  title = {{MiniLM}: Deep Self-Attention Distillation for Task-Agnostic Compression of Pre-Trained Transformers},
  booktitle = {Advances in Neural Information Processing Systems},
  volume = {33},
  year = {2020}
}

@inproceedings{wang2020uniformity,
  author = {Tongzhou Wang and Phillip Isola},
  title = {Understanding Contrastive Representation Learning through Alignment and Uniformity on the Hypersphere},
  booktitle = {Proceedings of the 37th International Conference on Machine Learning},
  pages = {9929--9939},
  year = {2020}
}

@inproceedings{zuo2026compression,
  author = {Chunsheng Zuo and Daniel Khashabi},
  title = {More Than Efficiency: Embedding Compression Improves Domain Adaptation in Dense Retrieval},
  booktitle = {Proceedings of the First Workshop on Structured Understanding, Retrieval, and Generation in the LLM Era},
  pages = {361--377},
  year = {2026}
}

@article{kwiatkowski2019natural,
  author = {Tom Kwiatkowski and Jennimaria Palomaki and Olivia Redfield and Michael Collins and Ankur P. Parikh and Chris Alberti and Danielle Epstein and Illia Polosukhin and Jacob Devlin and Kenton Lee and Kristina Toutanova and Llion Jones and Matthew Kelcey and Ming-Wei Chang and Andrew M. Dai and Jakob Uszkoreit and Quoc Le and Slav Petrov},
  title = {Natural Questions: A Benchmark for Question Answering Research},
  journal = {Transactions of the Association for Computational Linguistics},
  volume = {7},
  pages = {452--466},
  year = {2019},
  doi = {10.1162/tacl_a_00276}
}

@inproceedings{yang2018hotpotqa,
  author = {Zhilin Yang and Peng Qi and Saizheng Zhang and Yoshua Bengio and William W. Cohen and Ruslan Salakhutdinov and Christopher D. Manning},
  title = {{HotpotQA}: A Dataset for Diverse, Explainable Multi-hop Question Answering},
  booktitle = {Proceedings of the 2018 Conference on Empirical Methods in Natural Language Processing},
  pages = {2369--2380},
  year = {2018},
  doi = {10.18653/v1/D18-1259}
}

@inproceedings{thorne2018fever,
  author = {James Thorne and Andreas Vlachos and Christos Christodoulopoulos and Arpit Mittal},
  title = {{FEVER}: A Large-scale Dataset for Fact Extraction and Verification},
  booktitle = {Proceedings of the 2018 Conference of the North American Chapter of the Association for Computational Linguistics},
  pages = {809--819},
  year = {2018},
  doi = {10.18653/v1/N18-1074}
}

@inproceedings{hasibi2017dbpedia,
  author = {Faegheh Hasibi and Fedor Nikolaev and Chenyan Xiong and Krisztian Balog and Svein Erik Bratsberg and Alexander Kotov and Jamie Callan},
  title = {{DBpedia-Entity} v2: A Test Collection for Entity Search},
  booktitle = {Proceedings of the 40th International ACM SIGIR Conference on Research and Development in Information Retrieval},
  pages = {1265--1268},
  year = {2017},
  doi = {10.1145/3077136.3080751}
}

@article{voorhees2020treccovid,
  author = {Ellen M. Voorhees and Tasmeer Alam and Steven Bedrick and Dina Demner-Fushman and William R. Hersh and Kyle Lo and Kirk Roberts and Ian Soboroff and Lucy Lu Wang},
  title = {{TREC-COVID}: Constructing a Pandemic Information Retrieval Test Collection},
  journal = {SIGIR Forum},
  volume = {54},
  number = {1},
  pages = {1:1--1:12},
  year = {2020},
  doi = {10.1145/3451964.3451965}
}

@misc{iyer2017quora,
  author = {Shankar Iyer and Nikhil Dandekar and Kornel Csernai},
  title = {First Quora Dataset Release: Question Pairs},
  year = {2017},
  howpublished = {\url{https://data.quora.com/First-Quora-Dataset-Release-Question-Pairs}}
}

@article{diggelmann2020climate,
  author = {Thomas Diggelmann and Jordan L. Boyd-Graber and Jannis Bulian and Massimiliano Ciaramita and Markus Leippold},
  title = {{CLIMATE-FEVER}: A Dataset for Verification of Real-World Climate Claims},
  journal = {CoRR},
  volume = {abs/2012.00614},
  year = {2020}
}

@article{jarvelin2002cumulated,
  author = {Kalervo J{\"a}rvelin and Jaana Kek{\"a}l{\"a}inen},
  title = {Cumulated Gain-based Evaluation of {IR} Techniques},
  journal = {ACM Transactions on Information Systems},
  volume = {20},
  number = {4},
  pages = {422--446},
  year = {2002},
  doi = {10.1145/582415.582418}
}

@article{bengio2013estimating,
  author = {Yoshua Bengio and Nicholas L{\'e}onard and Aaron C. Courville},
  title = {Estimating or Propagating Gradients Through Stochastic Neurons for Conditional Computation},
  journal = {CoRR},
  volume = {abs/1308.3432},
  year = {2013}
}

@inproceedings{cao2017hashnet,
  author = {Zhangjie Cao and Mingsheng Long and Jianmin Wang and Philip S. Yu},
  title = {{HashNet}: Deep Learning to Hash by Continuation},
  booktitle = {Proceedings of the IEEE International Conference on Computer Vision},
  pages = {5609--5618},
  publisher = {IEEE Computer Society},
  year = {2017},
  doi = {10.1109/ICCV.2017.598}
}

@inproceedings{burges2005learning,
  author = {Christopher J. C. Burges and Tal Shaked and Erin Renshaw and Ari Lazier and Matt Deeds and Nicole Hamilton and Gregory N. Hullender},
  title = {Learning to Rank Using Gradient Descent},
  booktitle = {Proceedings of the Twenty-Second International Conference on Machine Learning},
  pages = {89--96},
  publisher = {Association for Computing Machinery},
  year = {2005},
  doi = {10.1145/1102351.1102363}
}

@article{hinton2015distilling,
  author = {Geoffrey E. Hinton and Oriol Vinyals and Jeffrey Dean},
  title = {Distilling the Knowledge in a Neural Network},
  journal = {CoRR},
  volume = {abs/1503.02531},
  year = {2015}
}

@inproceedings{qu2021rocketqa,
  author = {Yingqi Qu and Yuchen Ding and Jing Liu and Kai Liu and Ruiyang Ren and Wayne Xin Zhao and Daxiang Dong and Hua Wu and Haifeng Wang},
  title = {{RocketQA}: An Optimized Training Approach to Dense Passage Retrieval for Open-Domain Question Answering},
  booktitle = {Proceedings of the 2021 Conference of the North American Chapter of the Association for Computational Linguistics: Human Language Technologies},
  pages = {5835--5847},
  publisher = {Association for Computational Linguistics},
  year = {2021},
  doi = {10.18653/v1/2021.naacl-main.466}
}

@inproceedings{weiss2008spectral,
  author = {Yair Weiss and Antonio Torralba and Robert Fergus},
  title = {Spectral Hashing},
  booktitle = {Advances in Neural Information Processing Systems},
  volume = {21},
  pages = {1753--1760},
  year = {2008}
}

@article{robertson2009probabilistic,
  author = {Stephen E. Robertson and Hugo Zaragoza},
  title = {The Probabilistic Relevance Framework: {BM25} and Beyond},
  journal = {Foundations and Trends in Information Retrieval},
  volume = {3},
  number = {4},
  pages = {333--389},
  year = {2009},
  doi = {10.1561/1500000019}
}

@inproceedings{ji2012superbit,
  author = {Jianqiu Ji and Jianmin Li and Shuicheng Yan and Bo Zhang and Qi Tian},
  title = {Super-Bit Locality-Sensitive Hashing},
  booktitle = {Advances in Neural Information Processing Systems},
  volume = {25},
  year = {2012}
}

@inproceedings{covington2016youtube,
  author       = {Paul Covington and
                  Jay Adams and
                  Emre Sargin},
  title        = {Deep Neural Networks for YouTube Recommendations},
  booktitle    = {Proceedings of the 10th {ACM} Conference on Recommender Systems,
                  {RecSys} 2016, Boston, MA, USA, September 15-19, 2016},
  pages        = {191--198},
  publisher    = {{ACM}},
  year         = {2016},
  doi          = {10.1145/2959100.2959190}
}

@inproceedings{yi2019sampling,
  author       = {Xinyang Yi and
                  Ji Yang and
                  Lichan Hong and
                  Derek Zhiyuan Cheng and
                  Lukasz Heldt and
                  Aditee Kumthekar and
                  Zhe Zhao and
                  Li Wei and
                  Ed H. Chi},
  title        = {Sampling-bias-corrected neural modeling for large corpus item
                  recommendations},
  booktitle    = {Proceedings of the 13th {ACM} Conference on Recommender Systems,
                  {RecSys} 2019, Copenhagen, Denmark, September 16-20, 2019},
  pages        = {269--277},
  publisher    = {{ACM}},
  year         = {2019},
  doi          = {10.1145/3298689.3346996}
}

@inproceedings{wadden2020factfiction,
  author       = {David Wadden and
                  Shanchuan Lin and
                  Kyle Lo and
                  Lucy Lu Wang and
                  Madeleine van Zuylen and
                  Arman Cohan and
                  Hannaneh Hajishirzi},
  title        = {Fact or Fiction: Verifying Scientific Claims},
  booktitle    = {Proceedings of the 2020 Conference on Empirical Methods in Natural
                  Language Processing, {EMNLP} 2020, Online, November 16-20, 2020},
  pages        = {7534--7550},
  publisher    = {Association for Computational Linguistics},
  year         = {2020},
  doi          = {10.18653/v1/2020.emnlp-main.609}
}

@inproceedings{reimers2019sbert,
  author       = {Nils Reimers and
                  Iryna Gurevych},
  title        = {Sentence-{BERT}: Sentence Embeddings using Siamese {BERT}-Networks},
  booktitle    = {Proceedings of the 2019 Conference on Empirical Methods in Natural
                  Language Processing and the 9th International Joint Conference on
                  Natural Language Processing, {EMNLP-IJCNLP} 2019, Hong Kong, China,
                  November 3-7, 2019},
  pages        = {3982--3992},
  publisher    = {Association for Computational Linguistics},
  year         = {2019},
  doi          = {10.18653/v1/D19-1410}
}

@article{hua2026pointing,
  author       = {Peichun Hua and
                  Danyang Chen and
                  Junan Zhang and
                  Haifeng Sun and
                  Jingyu Wang and
                  Diwen Xue and
                  Mingyu Li and
                  Yunming Xiao},
  title        = {Pointing the Way, Hiding the Destination: Practical Private Dense Retrieval at Scale},
  journal      = {CoRR},
  volume       = {abs/2608.25735},
  year         = {2026},
  url          = {https://arxiv.org/abs/2608.25735},
  doi          = {10.48550/arXiv.2608.25735},
  eprinttype   = {arXiv},
  eprint       = {2608.25735}
}

@article{cui2026mess,
  author       = {Haoyu Cui and
                  Zengpeng Li and
                  Tien Tuan Anh Dinh and
                  Mei Wang},
  title        = {{MESS}: Fast and Private Semantic Search on Multi-Graph {HNSW}},
  journal      = {CoRR},
  volume       = {abs/2607.28999},
  year         = {2026},
  url          = {https://arxiv.org/abs/2607.28999},
  doi          = {10.48550/arXiv.2607.28999},
  eprinttype   = {arXiv},
  eprint       = {2607.28999},
  bibsource    = {dblp computer science bibliography, https://dblp.org}
}

@misc{hua2026spruce,
    title={Spruce: Scalable Private Outsourced Retrieval Using Compact Embeddings}, 
    author={Peichun Hua and Yunming Xiao},
    year={2026},
    eprint={2609.03376},
    archivePrefix={arXiv},
    primaryClass={cs.CR},
    url={https://arxiv.org/abs/2609.03376}, 
}

@article{nussbaum2025nomic,
  title   = {Nomic Embed: Training a Reproducible Long Context Text Embedder},
  author  = {Nussbaum, Zach and Morris, John X. and Duderstadt, Brandon and Mulyar, Andriy},
  journal = {Transactions on Machine Learning Research},
  year    = {2025}
}

@misc{koenig2024binarymrl,
  title        = {64 Bytes per Embedding, Yee-Haw},
  note         = {Binary MRL for mxbai-embed-large-v1},
  author       = {Koenig, Darius and Shakir, Aamir},
  year         = {2024},
  howpublished = {Mixedbread blog, \url{https://www.mixedbread.com/blog/binary-mrl}}
}

@article{nair2025matquant,
  title   = {Matryoshka Quantization},
  author  = {Nair, Pranav and Datta, Puranjay and Dean, Jeff and Jain, Prateek and Kusupati, Aditya},
  journal = {CoRR},
  volume  = {abs/2502.06786},
  year    = {2025}
}

@article{gao2025extendedrabitq,
  title   = {Practical and Asymptotically Optimal Quantization of High-Dimensional Vectors in {E}uclidean Space for Approximate Nearest Neighbor Search},
  author  = {Gao, Jianyang and Gou, Yutong and Xu, Yuexuan and Yang, Yongyi and Long, Cheng and Wong, Raymond Chi-Wing},
  journal = {Proceedings of the ACM on Management of Data},
  volume  = {3},
  number  = {3},
  pages   = {1--26},
  year    = {2025},
  doi     = {10.1145/3725413}
}

@misc{nomic2024v15,
  title        = {nomic-embed-text-v1.5 Model Card},
  author       = {{Nomic AI}},
  year         = {2024},
  howpublished = {\url{https://huggingface.co/nomic-ai/nomic-embed-text-v1.5}}
}

\appendix

\section{Seeds and Significance}
\label{app:seeds}

Table~\ref{tab:seeds} reports seed variation for the recipes compared in Figure~\ref{fig:pareto}. Stage-II seeds re-train only the adaptor on the fixed Stage-I model; direct nesting has no Stage II, so its seeds are full Stage-I runs of about 29 GPU-hours each. At 16 and 32 bytes, every Stage-II seed is above every direct-nesting seed. A second full-pipeline run, which also re-trains Stage I (seed 43, one block), reaches .365/.487/.547, matching direct nesting at 16 bytes and leading it by .009 at 32 bytes. The one-block adaptor of the conference version and its two learning rates are shown as well: the learning rate has no effect beyond seed variation, and the second block adds .005, .004, and .002 at 8, 16, and 32 bytes. Without TREC-COVID, whose 50 queries make it the noisiest dataset, MHR still exceeds direct nesting at 16 and 32 bytes (.477 against .473, and .530 against .520).

\begin{table}[ht]
\centering
{\small
\setlength{\tabcolsep}{3pt}
\begin{tabular}{@{}lcccc@{}}
\toprule
Recipe & Seeds & 8 B & 16 B & 32 B \\
\midrule
\textbf{MHR} (two blocks) & 3 & .370$\pm$.006 & .499$\pm$.003 & .553$\pm$.003 \\
MHR, one block & 5 & .365$\pm$.003 & .495$\pm$.004 & .551$\pm$.002 \\
\quad LR $5\!\times\!10^{-4}$ & 5 & .365$\pm$.003 & .493$\pm$.001 & .550$\pm$.003 \\
\quad Stage-I seed 43 & 1 & .365 & .487 & .547 \\
Direct nesting & 2 & .374$\pm$.007 & .487$\pm$.000 & .538$\pm$.001 \\
\bottomrule
\end{tabular}}
\caption{Seven-dataset macro NDCG@10, mean$\pm$standard deviation over seeds. MHR seeds are Stage-II seeds on one Stage-I model, except the Stage-I seed 43 row.}
\label{tab:seeds}
\end{table}

Against JPQ-FT, the paired bootstrap gives macro margins of +.159, +.096, and +.032 at 8, 16, and 32 bytes, and against JPQ-Nested +.188, +.147, and +.031, all with $p<.001$. At 32 bytes the 95\% interval against JPQ-Nested is [.022, .040].

\section{Fitting-Sample Variance}
\label{app:fitvar}

Every source-fitted baseline is fitted on a random 20K-passage MS~MARCO sample. We repeat each fit on five disjoint samples and report the mean in the main tables; Table~\ref{tab:fitvar} gives the spread. Macro NDCG@10 varies by at most .005, and TREC-COVID, with 50 queries, varies the most (up to .030). The same holds for the systems tables: OPQ FastScan and PQ flat vary by $\le$.002 on Quora and .030/.014 on TREC-COVID, IVF-PQ by $\le$.008 on Quora and .021--.036 on TREC-COVID, and LEANN PQ-S by $\le$.006. MHR is fitted once on the full source set and does not depend on these samples, apart from the shared IVF coarse quantizer.

\begin{table*}[t]
\centering
{\small
\setlength{\tabcolsep}{3.5pt}
\begin{tabular}{@{}lc*{8}{c}@{}}
\toprule
Method & B & Q & NQ & HP & DB & FV & TC & CF & Avg. \\
\midrule
PQ & 16 & .703{\tiny$\pm$.004} & .145{\tiny$\pm$.006} & .157{\tiny$\pm$.002} & .144{\tiny$\pm$.005} & .268{\tiny$\pm$.004} & .349{\tiny$\pm$.029} & .114{\tiny$\pm$.009} & .268{\tiny$\pm$.005} \\
OPQ & 16 & .757{\tiny$\pm$.004} & .253{\tiny$\pm$.006} & .245{\tiny$\pm$.004} & .215{\tiny$\pm$.003} & .424{\tiny$\pm$.005} & .448{\tiny$\pm$.023} & .146{\tiny$\pm$.019} & .355{\tiny$\pm$.004} \\
RaBitQ & 16 & .676{\tiny$\pm$.001} & .082{\tiny$\pm$.003} & .034{\tiny$\pm$.002} & .045{\tiny$\pm$.004} & .097{\tiny$\pm$.003} & .291{\tiny$\pm$.023} & .053{\tiny$\pm$.010} & .183{\tiny$\pm$.004} \\
PCA-RR & 16 & .773{\tiny$\pm$.002} & .207{\tiny$\pm$.003} & .153{\tiny$\pm$.003} & .143{\tiny$\pm$.007} & .308{\tiny$\pm$.007} & .450{\tiny$\pm$.015} & .085{\tiny$\pm$.010} & .303{\tiny$\pm$.002} \\
ITQ & 16 & .772{\tiny$\pm$.003} & .203{\tiny$\pm$.005} & .147{\tiny$\pm$.001} & .149{\tiny$\pm$.004} & .312{\tiny$\pm$.003} & .458{\tiny$\pm$.023} & .098{\tiny$\pm$.019} & .305{\tiny$\pm$.005} \\
\midrule
PQ & 32 & .828{\tiny$\pm$.002} & .363{\tiny$\pm$.003} & .468{\tiny$\pm$.003} & .295{\tiny$\pm$.007} & .638{\tiny$\pm$.006} & .559{\tiny$\pm$.014} & .190{\tiny$\pm$.009} & .478{\tiny$\pm$.004} \\
OPQ & 32 & .832{\tiny$\pm$.002} & .402{\tiny$\pm$.005} & .497{\tiny$\pm$.003} & .328{\tiny$\pm$.005} & .680{\tiny$\pm$.008} & .602{\tiny$\pm$.020} & .206{\tiny$\pm$.013} & .507{\tiny$\pm$.004} \\
RaBitQ & 32 & .848{\tiny$\pm$.001} & .391{\tiny$\pm$.003} & .424{\tiny$\pm$.004} & .296{\tiny$\pm$.005} & .655{\tiny$\pm$.008} & .637{\tiny$\pm$.030} & .194{\tiny$\pm$.006} & .492{\tiny$\pm$.005} \\
PCA-RR & 32 & .840{\tiny$\pm$.001} & .355{\tiny$\pm$.003} & .395{\tiny$\pm$.002} & .264{\tiny$\pm$.004} & .610{\tiny$\pm$.005} & .604{\tiny$\pm$.030} & .166{\tiny$\pm$.012} & .462{\tiny$\pm$.004} \\
ITQ & 32 & .840{\tiny$\pm$.001} & .362{\tiny$\pm$.006} & .390{\tiny$\pm$.003} & .261{\tiny$\pm$.004} & .605{\tiny$\pm$.005} & .612{\tiny$\pm$.017} & .168{\tiny$\pm$.009} & .462{\tiny$\pm$.003} \\
\bottomrule
\end{tabular}}
\caption{Source-fitted baselines over five MS~MARCO fitting samples: NDCG@10 mean$\pm$std.}
\label{tab:fitvar}
\end{table*}

\section{More Datasets and In-Domain Quality}
\label{app:datasets}

Table~\ref{tab:beir13} adds the six smaller public BEIR datasets (ArguAna, FiQA, NFCorpus, SciDocs, SciFact, Touch\'e-2020), giving all 13 public sets, and the in-domain MS~MARCO dev set (6,980 queries). On the 13 sets, MHR has the best macro NDCG@10 at every budget. Per dataset (paired bootstrap, 95\%), it is significantly better than JPQ-FT on 12 of 13 sets at 8 and 16 bytes; at 32 bytes it wins 6, loses 2, and ties 5. Against JPQ-Nested the counts are 12/0/1, 12/0/1, and 5/1/7. Table~\ref{tab:beir6} shows the two 32-byte losses: ArguAna and Touch\'e-2020, the latter with 49 queries.

In domain, MHR leads JPQ-FT by .103 and .033 at 8 and 16 bytes and stays within .006 at 32 bytes, although JPQ-FT fine-tunes the whole encoder on MS~MARCO. Direct nesting is slightly ahead in domain: the full-width anchor gives up this small source-domain edge for the transfer gains of Table~\ref{tab:dataset}.

\begin{table*}[ht]
\centering
{\small
\setlength{\tabcolsep}{4pt}
\begin{tabular}{@{}lccccccccc@{}}
\toprule
& \multicolumn{3}{c}{All 13 BEIR sets} & \multicolumn{3}{c}{Six added sets} & \multicolumn{3}{c}{MS~MARCO dev} \\
\cmidrule(lr){2-4}\cmidrule(lr){5-7}\cmidrule(lr){8-10}
Method & 8 B & 16 B & 32 B & 8 B & 16 B & 32 B & 8 B & 16 B & 32 B \\
\midrule
\textbf{MHR} & \textbf{.315} & \textbf{.412} & \textbf{.458} & .244 & .308 & .343 & .256 & .320 & .349 \\
Direct nesting & .312 & .405 & .446 & \textbf{.246} & \textbf{.309} & .337 & \textbf{.262} & \textbf{.322} & .353 \\
JPQ-FT & .198 & .346 & .445 & .175 & .275 & .352 & .153 & .287 & \textbf{.355} \\
JPQ-Nested & .167 & .302 & .447 & .143 & .242 & \textbf{.355} & .137 & .259 & .350 \\
\midrule
Float BGE & \multicolumn{6}{c}{} & \multicolumn{3}{c}{.414} \\
\bottomrule
\end{tabular}}
\caption{Macro NDCG@10 on all 13 public BEIR sets, the six sets beyond Table~\ref{tab:budget}, and the in-domain MS~MARCO dev set. Bold: best compact method per column.}
\label{tab:beir13}
\end{table*}

\begin{table}[ht]
\centering
{\small
\setlength{\tabcolsep}{3pt}
\begin{tabular}{@{}lrcccccc@{}}
\toprule
& & \multicolumn{3}{c}{16 B} & \multicolumn{3}{c}{32 B} \\
\cmidrule(lr){3-5}\cmidrule(lr){6-8}
Dataset & $n_q$ & MHR & FT & N & MHR & FT & N \\
\midrule
ArguAna & 1406 & \textbf{.363} & .333 & .299 & .407 & \textbf{.425} & .423 \\
FiQA & 648 & \textbf{.247} & .223 & .184 & .306 & \textbf{.312} & .308 \\
NFCorpus & 323 & \textbf{.311} & .266 & .227 & \textbf{.338} & .321 & .326 \\
SciDocs & 1000 & \textbf{.158} & .126 & .116 & \textbf{.187} & .175 & .177 \\
SciFact & 300 & \textbf{.610} & .528 & .466 & .658 & .649 & \textbf{.667} \\
Touch\'e-2020 & 49 & .160 & \textbf{.177} & .159 & .164 & \textbf{.230} & .227 \\
\bottomrule
\end{tabular}}
\caption{NDCG@10 on the six added BEIR sets. FT: JPQ-FT; N: JPQ-Nested; $n_q$: test queries.}
\label{tab:beir6}
\end{table}

\section{Binary MRL and Backbone Dependence}
\label{app:mrl}

Binary MRL signs 64/128/256-dimension prefixes of the off-the-shelf Matryoshka encoders nomic-embed-v1.5 and mxbai-embed-large \cite{nussbaum2025nomic,koenig2024binarymrl}. Table~\ref{tab:mrl} compares it with MHR. Asymmetric scoring is stronger than Hamming scoring for every encoder. Both MRL encoders are stronger than BGE in float, yet at 32 bytes they retain 79\% and 80\% of their float quality, against 87\% for MHR. Truncating and signing BGE itself, which was not trained for nesting, gives .157/.323/.488.

Subtracting the mean of MS~MARCO documents before signing strengthens every encoder. Centered mxbai (.245/.415/.533) and centered BGE (.188/.351/.504) stay below MHR at every budget. Centered nomic reaches .286/.460/.568, .012 above MHR on BGE at 32 bytes, with its gain concentrated on NQ, HotpotQA, and FEVER, whose training splits nomic-embed uses in supervised fine-tuning \cite{nussbaum2025nomic,nomic2024v15}: on the ten public BEIR sets that neither model trained on, MHR leads at every budget (.307/.385/.422 against .240/.343/.407).

To separate method from backbone, we trained MHR on the nomic encoder (one-block Stage II on a 12-epoch Stage I, half the BGE schedule). It reaches .345/.497/.570 and leads centered nomic at every budget, by .059 and .037 at 8 and 16 bytes ($p<.001$) and .002 at 32 bytes; against plain binary MRL the margins are .14, .11, and .05.

\begin{table}[ht]
\centering
{\small
\setlength{\tabcolsep}{3pt}
\begin{tabular}{@{}lcccc@{}}
\toprule
Encoder and scoring & Float & 8 B & 16 B & 32 B \\
\midrule
nomic, Hamming & .662 & .114 & .253 & .422 \\
nomic, asymmetric & .662 & .205 & .384 & .521 \\
nomic, centered asym. & .662 & .286 & .460 & .568 \\
mxbai, Hamming & .655 & .142 & .282 & .439 \\
mxbai, asymmetric & .655 & .206 & .394 & .526 \\
mxbai, centered asym. & .655 & .245 & .415 & .533 \\
BGE, asymmetric & .637 & .157 & .323 & .488 \\
BGE, centered asym. & .637 & .188 & .351 & .504 \\
\midrule
MHR on nomic & .662 & .345 & .497 & .570 \\
\textbf{MHR} on BGE & .637 & .376 & .502 & .556 \\
\bottomrule
\end{tabular}}
\caption{Binary Matryoshka baselines, seven-dataset macro NDCG@10. Float is the encoder's full-dimension float quality.}
\label{tab:mrl}
\end{table}

\paragraph{In-distribution supplement.} We also trained Stage II on the official NQ, HotpotQA, and FEVER training splits, the data nomic fine-tunes on, keeping Stage I unchanged. On BGE (one-block) this raises 8 bytes by .014 ($p<.001$) and keeps 16 and 32 bytes (.376/.497/.553). On the three in-distribution datasets it reaches .367/.513/.582, against .239/.483/.619 for centered nomic. On the nomic backbone, the same supplement gives .380/.512/.568 over seven datasets, .094 and .052 above centered nomic at 8 and 16 bytes.

\section{Controlled MRL Training}
\label{app:mrlctrl}

The off-the-shelf comparison mixes method and training data. As a controlled counterpart, we trained Stage I with MRL-style nested supervision on relaxed codes, with our backbone, data, and asymmetric evaluation, and signed its prefixes. It reaches .380/.491/.545: ahead at 8 bytes, and behind MHR at 16 and 32 bytes (.502/.556). Running our Stage II on top of it gives .380/.494/.546, since its Stage I already produces nested prefixes. MHR obtains its 16- and 32-byte quality post hoc, from an existing single-width code and its index, with a frozen encoder and about 2 GPU-hours of Stage II, against about 29 GPU-hours for a full MRL-style encoder run.

\section{Sensitivity}
\label{app:sens}

Table~\ref{tab:sens} varies every Stage-II setting around the one-block recipe, each with one seed. All runs trained stably, and no setting is significantly better than the default under the paired bootstrap. Halving or doubling the anchor and relevance weights moves macro NDCG@10 by at most .014, and the 32-byte score moves by at most .005 for every setting. The prefix-weight vector, adaptor width, and anchor temperature have small effects. Removing the anchor (Table~\ref{tab:ablations}) is the only large change. Training only the 64- and 256-bit widths still gives .488 at the untrained 128-bit prefix, so the nested organization extends to widths without a direct loss, and adding a 32-bit width (.163 at 4 bytes) costs .011 at 8 bytes. Training the widths one after another instead of jointly costs .014 at 16 bytes.

\begin{table}[ht]
\centering
{\small
\setlength{\tabcolsep}{2.5pt}
\begin{tabular}{@{}lccc@{}}
\toprule
Setting & 8 B & 16 B & 32 B \\
\midrule
Default (one block) & .363 & .500 & .552 \\
\midrule
$\alpha_{64}=0.5$ & .362 & .495 & .552 \\
$\alpha_{64}=1.5$ & .366 & .491$^\dagger$ & .547$^\dagger$ \\
$\alpha_{64}=2$ & .365 & .492$^\dagger$ & .553 \\
uniform $\alpha$ & .364 & .501 & .554 \\
$\lambda_{\mathrm{src}}\times0.5$ / $\times2$ & .365 / .349$^\dagger$ & .493$^\dagger$ / .488$^\dagger$ & .550 / .550 \\
$\lambda_{\mathrm{rel}}\times0.5$ / $\times2$ & .350$^\dagger$ / .367 & .488$^\dagger$ / .488$^\dagger$ & .550 / .550 \\
$\lambda_{\mathrm{ce}}\times0$ / $\times2$ & .359 / .367 & .497 / .499 & .553 / .552 \\
$\lambda_{\mathrm{gor}}\times0$ & .366 & .496$^\dagger$ & .552 \\
$\lambda_{\mathrm{bal}}\times0$ & .363 & .495$^\dagger$ & .552 \\
anchor temperature $\times2$ & .370 & .496 & .549 \\
hidden width 128 / 512 & .365 / .364 & .495 / .498 & .552 / .550 \\
widths $\{64,256\}$ & .363 & .488$^*$ & .550 \\
widths $\{32,64,128,256\}$ & .351$^\dagger$ & .495 & .549 \\
sequential widths & .367 & .486$^\dagger$ & .550 \\
\bottomrule
\end{tabular}}
\caption{Stage-II sensitivity, seven-dataset macro NDCG@10. $^\dagger$: significantly below the default (paired bootstrap, 95\%). $^*$: width not trained.}
\label{tab:sens}
\end{table}



\section{Prefix Cascade on One Stored Code}
\label{app:cascade}

Because every MHR prefix is itself a valid binary code, a search can scan only a prefix of the stored 32-byte code and re-score a shortlist with the full asymmetric score, with no extra storage. PQ codes are not nested, so a PQ index would need a second stored code for the same cascade. Table~\ref{tab:systems}(a) reports the 16-byte variant: symmetric Hamming over the prefix using AVX-512 population counts, selection of the $K'$ nearest documents, and exact asymmetric re-scoring over all 256 bits, implemented as a single-thread C kernel. The stored code is laid out as the prefix words in column order followed by the row-major remainder, still 32 bytes per document, and the kernel scans the prefix for blocks of 16 queries at a time. We report the smallest shortlist within .005 NDCG@10 of scanning the full code: $K'=200$ on Quora and $K'=2000$ on TREC-COVID give .8495/.7025 at .132/.118 ms per query, against .8501/.7075 at .490/.192 ms for full-code FastScan. Larger shortlists close the remaining TREC-COVID gap (.704 at $K'=5000$, .196 ms), and an 8-byte prefix is faster still on Quora (.845 at .099 ms with $K'=500$). All timings are single-thread on one core, measured like Table~\ref{tab:systems}(a), and all FastScan rows use the fastest FAISS FastScan variant (\texttt{implem} 15). At 16 bytes under the same FastScan kernel, the MHR prefix reaches .827/.641 NDCG@10 at .307/.130 ms on Quora/TREC-COVID, against .768/.453 (mean of five fitting samples) at .320/.143 ms for source-fitted OPQ, which also stores 20.6/30.1 B/doc including its rotation. The same cascade within the 16-byte code, an 8-byte popcount prefix with exact 16-byte re-scoring of $K'=500$/$1000$ candidates, reaches .824/.651 at .099/.073 ms, 3.2$\times$/2.0$\times$ faster than 16-byte OPQ FastScan and more accurate.

\end{document}